\documentclass[%
  journal=jctcce
]{achemso}

\usepackage[utf8]{inputenc}
\usepackage[T1]{fontenc}
\usepackage{xcolor}
\definecolor{blue}{RGB}{0,0,153}
\usepackage[%
  colorlinks=true,
  allcolors=blue
]{hyperref}
\usepackage{url}
\usepackage{enumitem}
\usepackage{amsfonts}
\usepackage{amssymb}
\usepackage{amsmath}
\usepackage{mathtools}
\usepackage[ruled,vlined]{algorithm2e}
\usepackage{lmodern}
\usepackage[normalem]{ulem}

\DeclarePairedDelimiter\ppar{(}{)}              
\DeclarePairedDelimiter\pang{\langle}{\rangle}  
\DeclarePairedDelimiter\pnrm{\lVert}{\rVert}    
\DeclarePairedDelimiter\pbkt{[}{]}              
\DeclarePairedDelimiter\pset{\{}{\}}            

\newcommand{\rfig}[1]{Fig.~\ref{#1}}

\newcommand{\rsct}[1]{Sec.~\ref{#1}}
\newcommand{\rref}[1]{Ref.~\citenum{#1}}
\newcommand{\req}[1]{Eq.~\ref{#1}}

\newcommand{\pr}{\operatorname{Pr}\,}
\newcommand{\dd}[1]{\operatorname{d#1}}

\newcommand{\bz}{\mathbf{z}}
\newcommand{\bx}{\mathbf{x}}
\newcommand{\bs}{\mathbf{s}}

\newcommand{\dz}{\dd{\mathbf{z}}}
\newcommand{\ds}{\dd{\mathbf{s}}}
\newcommand{\dx}{\dd{\mathbf{x}}}
\newcommand{\e}{\operatorname{e}}
\newcommand{\kT}{k_{\mathrm{B}}T}

\newcommand{\cond}{\left.\right\vert}

\newcommand{\chg}[1]{{\color{black}{#1}}}

\title{\Large Spectral Map for Slow Collective Variables, Markovian Dynamics, and Transition State Ensembles}

\author{Jakub Rydzewski}
\email{jr@fizyka.umk.pl}
\affiliation{%
  Institute of Physics,
  Faculty of Physics, Astronomy and Informatics,
  Nicolaus Copernicus University,
  Grudziadzka 5, 87-100 Toru\'n, Poland
}

\begin{document}

\begin{tocentry}
  \begin{center}
    \includegraphics{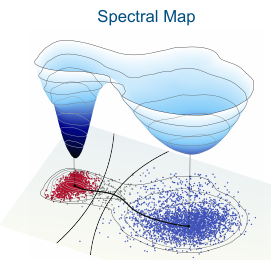}
  \end{center}
\end{tocentry}

\newpage

\begin{abstract}
Understanding the behavior of complex molecular systems is a fundamental problem in physical chemistry. To describe the long-time dynamics of such systems, which is responsible for their most informative characteristics, we can identify a few slow collective variables (CVs) while treating the remaining fast variables as thermal noise. This enables us to simplify the dynamics and treat it as diffusion in a free-energy landscape spanned by slow CVs, effectively rendering the dynamics Markovian. Our recent statistical learning technique, spectral map \href{https://doi.org/10.1021/acs.jpclett.3c01101}{[Rydzewski, {\it J. Phys. Chem. Lett.} {\bf 2023}, 14, 22, 5216--5220]}, explores this strategy to learn slow CVs by maximizing a spectral gap of a transition matrix. In this work, we introduce several advancements into our framework, using a high-dimensional reversible folding process of a protein as an example. We implement an algorithm for coarse-graining Markov transition matrices to partition the reduced space of slow CVs kinetically and use it to define a transition state ensemble. We show that slow CVs learned by spectral map closely approach the Markovian limit for an overdamped diffusion. We demonstrate that coordinate-dependent diffusion coefficients only slightly affect the constructed free-energy landscapes. Finally, we present how spectral map can be used to quantify the importance of features and compare slow CVs with structural descriptors commonly used in protein folding. \chg{Overall, we demonstrate that a single slow CV learned by spectral map can be used as a physical reaction coordinate to capture essential characteristics of protein folding.}
\end{abstract}

\maketitle

\newpage

\section{Introduction}
\chg{Learning the complex behavior of systems with multiple timescales poses a fundamental problem in physical chemistry and molecular dynamics simulations~\cite{bolhuis2002transition,valsson2016enhancing,roux2022transition}.} Such systems often display slow dynamics toward equilibrium due to rare transitions between long-lived metastable states while also retaining fast fluctuations within these states. The infrequent transitions occur near free-energy barriers much higher than the thermal energy ($\gg \kT$) and form a transition state ensemble. Together, these processes give rise to an event known as timescale separation, which is often described as a hallmark of barrier-crossing dynamics. To simplify the representation of such complex systems, a few functions of microscopic coordinates are often introduced, referred to as collective variables (CVs). However, accurately capturing these characteristics in the reduced representation, especially transition state ensembles~\cite{best2005reaction,lechner2010nonlinear,wang2021state,he2022committor,ray2023deep}, remains a persistent issue~\cite{peters2016reaction}. 

\chg{Relying only on intuition or trial and error to identify CVs can be unsystematic and obscure our understanding of the underlying physical process, contributing to erroneously estimated kinetics. This can often manifest as:}
\begin{enumerate}
  \item Overlapping metastable states, which results in the underestimation of free-energy barriers, inaccurate determination of transition state ensembles, and inefficiency of enhanced sampling techniques due to the existence of hidden bottlenecks~\cite{valsson2016enhancing,peters2013reaction}.
  \item Inability to extract the behavior of the process on longer timescales (e.g., mixing slow and fast variables), and thus considerable non-Markovian effects~\cite{kneller2001computing,min2005observation,lange2006collective,micheletti2008optimal} that should then be additionally accounted for using a generalized Langevin equation with a memory kernel as in the Mori--Zwanzig formalism~\cite{zwanzig1961memory,mori1965transport,zwanzig2001nonequilibrium}.
\end{enumerate}
To alleviate these problems, many methods for the determination of CVs have recently been developed at the intersection of statistical physics, machine learning, and molecular dynamics. Some notable techniques construct CVs based on timescale separation~\cite{NIPS2005_2a0f97f8,singer2009detecting,tiwary2016spectral,chiavazzo2017intrinsic,mccarty2017variational,hovan2018defining,yang2018refining,wehmeyer2018time,chen2019nonlinear,bonati2021deep,novelli2022characterizing,chen2023discovering,shmilovich2023girsanov}, committors or splitting probabilities~\cite{hummer2004transition,ma2005automatic,metzner2009transition,lechner2010nonlinear,peters2013reaction,li2019computing,he2022committor,lazzeri2023molecular,kang2024computing}, or transfer operators~\cite{zhang2016effective,mcgibbon2017identification,wu2017variational,klus2018data,thiede2019galerkin,lorpaiboon2024accurate}. For more examples, see recent reviews~\cite{rohrdanz2013discovering,li2014recent,peters2016reaction,noe2017collective,wang2020machine,chen2021collective,chen2023chasing,rydzewski2023manifold,mehdi2024enhanced} and references therein.

In this work, we consider the construction of CVs for a complex system from the perspective of the spectral definition of metastability in a setting of Markov processes~\cite{gaveau1996master,gaveau1998theory,bovier2002metastability}, where rare transitions between metastable states can be related to the slow dynamics of CVs and timescale separation. This assumption allows for a description of the behavior of the system without dependence on previous states, making the reduced dynamics memoryless~\cite{guttenberg2013minimizing}. Then, the long-timescale dynamics is effectively Markovian and can be completely described by a free-energy landscape and diffusion coefficients~\cite{berezhkovskii2005one,berezhkovskii2011time}, without calculating complicated memory terms. Under this view, the dynamics primarily depends on slowly varying variables $\bz$, i.e., variables along which the system relaxes much more slowly than for any other variable, $\ll\tau_{\bz}$~\cite{du1998transition}. The fast variables adiabatically equilibrate to the slow variables and are treated as uncoupled thermal noise that introduces additional friction~\cite{berezhkovskii2011time}. Such a low-dimensional diffusive description has been useful in many processes~\cite{zwanzig1990rate}.

This is the direction taken by our recent unsupervised statistical learning technique called spectral map~\cite{rydzewski2023spectral}, developed based on parametric dimensionality reduction~\cite{hinton2006reducing,rydzewski2021multiscale,rydzewski2022reweighted} and the spectral theory of Markov processes~\cite{gaveau1996master,gaveau1998theory,bovier2002metastability,belkin2003laplacian,coifman2006diffusion,coifman2008diffusion}. Spectral map learns Markovian dynamics in the reduced space given by slow CVs by maximizing a spectral gap between slow and fast eigenvalues of a Markov transition matrix, increasing timescale separation and minimizing large memory effects. \chg{Our learning algorithm estimates transition probabilities adaptively based on an anisotropic diffusion kernel. This kernel encodes the geometry and density of data in reduced space, allowing for the representation of multiscale and heterogeneous free-energy landscapes with long-lived metastable states~\cite{rydzewski2024learning}.}

As a high-dimensional and non-trivial example of a molecular process, we choose the paradigmatic problem of protein folding. We employ spectral map framework to construct a one-dimensional slow CV for the reversible folding process of the FiP35 protein in solvent~\cite{shaw2010atomic}. FiP35 can be considered as a building block for understanding more complex proteins. We introduce an algorithm for kinetic partitioning of the CV space, which allows us to learn a transition state ensemble. Next, we use a test derived based on the transition state theory~\cite{berezhkovskii2018single} to show to what degree the slow CV learned by spectral map is Markovian. Moreover, we inspect how coordinate-dependent diffusion coefficients affect the free-energy profile along the slow CV. Finally, we show how to use spectral map as a feature selection pipeline and qualitatively compare the slow CV with frequent structural descriptors for protein folding.

\section{Framework}
\label{sec:framework}
\subsection{Reduced Space}
\label{sec:cv}
Consider a high-dimensional system described by $n$ configuration variables (i.e., features) $\bx = (x_1, \dots, x_n)$ whose dynamics at temperature $T$ under a potential energy function $U(\bx)$ is sampled from an unknown equilibrium distribution. Suppose we represent the system by its microscopic coordinates. In that case, the dynamics follows a canonical equilibrium distribution given by the Boltzmann density $p(\bx) = \e^{-\beta U(\bx)}/Z_U$, where $\beta=1/(k_{\mathrm{B}}T)$ is the inverse temperature, $k_{\mathrm{B}}$ is the Boltzmann constant and $Z_U=\int\dx\e^{-\beta U(\bx)}$ is the partition function of the system.

We map the high-dimensional configuration space into a reduced space $\bz=\ppar*{z_1, \dots, z_d}$ given by a set of $d$ functions of the configuration variables commonly referred to as CVs, where $d \ll n$. We encapsulate these functions in a target mapping~\cite{rydzewski2021multiscale,rydzewski2022reweighted,rydzewski2023manifold}:
\begin{equation}
  \label{eq:target-mapping}
  \bz = \xi_{w}(\bx) \equiv \pset[\big]{\xi_k(\bx, w)}_{k=1}^d,
\end{equation}
where $w$ are parameters ensuring that the target mapping describes the dynamics accurately. By sampling the system in the reduced space, its dynamics proceeds under a free-energy landscape (i.e., a potential of mean force):
\begin{align}
  \label{eq:fel}
  F(\bz) &= -\frac{1}{\beta}\log
    \int\dx\,\delta\ppar*{\bz - \xi_w(\bx)}
    \e^{-\beta U(\bx)}
\end{align}
so that the corresponding marginal equilibrium density is $p(\bz)=\e^{-\beta F(\bz)}/Z_F$, where $Z_F=\int\dz\e^{-\beta F(\bz)}$ is the reduced partition function.

\subsection{Fokker--Planck Diffusion}
\label{sec:fpd}
We assume that Markovian dynamics along slow coordinates can be represented as a diffusion process in the free-energy landscape. In this view, the reduced dynamics of the system can be described by an overdamped Langevin equation~\cite{berezhkovskii2011time,lu2014exact}:
\begin{equation}
  \label{eq:langevin}
  \dot{\bz} =  -\beta D(\bz) \nabla F(\bz) + 
    \nabla\cdot D(\bz) + 
    \sqrt{2 D(\bz)} \eta(t),
\end{equation}
where $D(\bz)$ is the coordinate-dependent diffusion tensor~\cite{hummer2005position,maragliano2006string} and $\eta(t)$ is a $d$-dimensional white-noise process satisfying $\pang{\eta_k(t)}=0$ and $\pang{\eta_k(t)\eta_l(s)}=\delta_{kl}\delta(t-s)$. As in \rref{nadler2006diffusion}, we simplify \req{eq:langevin} by setting the diffusion tensor $D(\bz)$ to unity. With the reduced dynamics given by the overdamped Langevin equation, the probability $p(\bz)$ satisfies the following forward Fokker--Planck (or Smoluchowski) equation:
\begin{equation}
  \label{eq:smoluchowski}
  \frac{\partial p(\bz)}{\partial t} = 
    \nabla \e^{-\beta F(\bz)} 
    \nabla \e^{\beta F(\bz)} p(\bz) \equiv 
    \operatorname{\mathcal{L}}_f p(\bz),
\end{equation}
which describes the time-propagation of the probability density $p(\bz)$ with the related generator of the diffusion process $\operatorname{\mathcal{L}}_f$. \chg{Alternatively, we can use a generator that corresponds to the backward Fokker--Planck equation:
\begin{equation}
  \label{eq:backward}
  \operatorname{\mathcal{L}}_b = \e^{\beta F(\bz)} 
    \nabla \e^{-\beta F(\bz)} 
    \nabla,
\end{equation}
that describes the evolution of $\bz$ in time. This pair of operators is adjoint and thus obeys the identity $\pang{g\cond \operatorname{\mathcal{L}}_f h}=\pang{\operatorname{\mathcal{L}}_b g\cond h}$ with the standard inner product.} Under general conditions, the generator of the diffusion process $\operatorname{\mathcal{L}}_f$ has a discrete eigenspectrum of non-positive eigenvalues $\pset{\mu_k}$ with $\mu_0 = 0 \ge \mu_1 \geq \mu_2 \geq \dots \geq \mu_{\infty}$, and corresponding eigenvectors $\psi_k(\bz)$. The general solution of \req{eq:smoluchowski} can be written in closed form as:
\begin{equation}
  \label{eq:prob}
  p(\bz) = \sum_{k=0}^{\infty} a_k \e^{-\mu_k t} \psi_k(\bz),
\end{equation}
where $a_k$ are coefficients. The long-time dynamics of the probability $p(\bz)$ converges the Boltzmann density in the free-energy landscape:
\begin{equation}
  \lim_{t\rightarrow\infty} p(\bz) = \psi_0(\bz) \propto 
    \e^{-\beta F(\bz)}.
\end{equation}
For the system with timescale separation, only a few slow processes corresponding to rare transitions between metastable states remain, so the eigenspectrum of the generator $\mathcal{L}_f$ has a spectral gap (also referred to as ``eigengap'') between the eigenvalues, $\mu_{l-1} \gg \mu_{l}$. 

This is due to the relation between the spectral gap (and thus the degree of degeneracy in the eigenvalue spectrum) and timescale separation~\cite{bovier2002metastability,gaveau1996master,gaveau1998theory}. Namely, if an eigenvalue is nearly degenerate $l$ times, it indicates that the equilibrium distribution breaks into $l-1$ metastable states with infrequent transitions between them. The converse is also true: if the equilibrium density breaks into metastable states separated by a free-energy barrier much larger than the thermal energy, there is eigenvalue degeneracy. Thus, the long-term dynamics of the system can be approximated by:
\begin{equation}
  \label{eq:prob-red}
  p(\bz) \approx \psi_0(\bz) + 
    \sum_{k=1}^{l-1} a_k \e^{-\mu_k t} \psi_k(\bz).
\end{equation}

\subsection{Markov Operator}
\label{sec:mto}
The Fokker--Planck stochastic diffusion given by the Smoluchowski equation (\req{eq:langevin}) is associated with a forward Markov operator $\operatorname{\mathcal{M}}_f$, which can be constructed using kernels. We consider an anisotropic diffusion kernel~\cite{coifman2005geometric,coifman2006diffusion} in the reduced space:
\begin{equation}
  \label{eq:aniso-kernel}
  a_{\alpha}(\bz,\bs) = 
    \frac{g(\bz,\bs)}{\varrho^{\alpha}(\bz)\varrho^{\alpha}(\bs)},
  \footnote{The term ``anisotropic diffusion'' should not be confused with a kernel that accounts for the coordinate-dependent diffusion tensor $D(\bz)$ (\req{eq:langevin}).}
\end{equation}
where $g(\bz,\bs)=\exp\ppar*{-{\pnrm{\bz-\bs}^2/\varepsilon}}$ is a Gaussian kernel with a scale constant $\varepsilon$, $\varrho(\bz)=\int\ds g(\bz,\bs) p(\bs)$ is a kernel density estimate, and $\alpha\in[0,1]$ is an anisotropic diffusion constant. Then, the anisotropic diffusion kernel is row-normalized by:
\begin{equation}
  d_{\alpha}(\bz) = \int\ds a_{\alpha}(\bz,\bs) p(\bs)
\end{equation}
to define the forward Markov transition kernel:
\begin{equation}
  \label{eq:mzs}
  M(\bz \cond \bs) = \pr \pset[\bbig]{\bz_{t+1} = 
    \bz \cond \bz_{t} = \bs} = 
      \frac{a_{\alpha}(\bz,\bs)}{d_{\alpha}(\bs)},
\end{equation}
where $M(\bz \cond \bs)$ is a probability that the system transitions to $\bz$ given that it is currently in $\bs$ after a time step in an auxiliary time $t$.

To establish the connection to the Fokker--Planck diffusion equation (\req{eq:langevin}), we first consider the forward Markov operator given by the following Chapman--Kolmogorov equation~\cite{risken1996fokker}:
\begin{equation}
  \operatorname{\mathcal{M}}_f\pbkt{f}(\bz) = 
    \int\ds M(\bz \cond \bs) f(\bs) p(\bs),
\end{equation}
acting on a dummy function $f$. Then, the generator of the Fokker--Planck diffusion $\mathcal{L}_f$ is related to the forward Markov operator by:
\begin{equation}
  \mathcal{L}_f = \lim_{\varepsilon\rightarrow 0} 
    \frac{\operatorname{\mathcal{M}}_f - \mathcal{I}}{\varepsilon},
\end{equation}
where $\mathcal{I}$ is an identity operator. Similarly, we can relate the generator $\mathcal{L}_b$ to the backward Markov operator $\operatorname{\mathcal{M}}_b\pbkt{g}$ constructed from the kernel $M(\bs\cond\bz)$.

We can see that the eigenvalues of the forward Markov operator are related to the eigenvalues of the forward Fokker--Plank diffusion equation by $\mu_k \approx (\lambda_k - 1)/\varepsilon$. That is, the eigendecomposition of the forward Markov operator yields the non-negative eigenvalues $\pset{\lambda_k}$, $\lambda_0 = 1 > \lambda_1 \dots \ge \lambda_\infty$ and eigenfunctions $f_k(\bz)$, where the eigenvalue $\lambda_0$ corresponds to the Boltzmann equilibrium distribution given by the eigenvector $f_0 \propto \e^{-\beta F(\bz)}$.

The anisotropic diffusion constant $\alpha$ introduced in \req{eq:aniso-kernel} can be used to define a class of anisotropic diffusion kernels with different limiting diffusion processes ($\varepsilon\rightarrow 0$)~\cite{coifman2005geometric,coifman2006diffusion}:
\begin{enumerate}
  \item For $\alpha=0$, this construction yields the classical normalized graph Laplacian with the generator corresponding to the backward Fokker--Planck equation with a free energy $2F(\bz)$.
  \item For $\alpha=1$, the backward generator gives the Laplace--Beltrami operator with the uniform probability density in the reduced space.
  \item For $\alpha=1/2$, the generator of the forward and backward operators coincide and correspond to the backward Fokker--Planck equation with a free energy $F(\bz)$.
\end{enumerate}
Therefore, the case with $\alpha = 1/2$ provides a consistent method to approximate the eigenvalues and eigenfunctions corresponding to the stochastic differential equation, which gives in the asymptotic limit the Fokker--Planck diffusion under the free-energy landscape with the marginal density $p(\bz)$ approaching the Boltzmann equilibrium density.

For convenience, we prefer to work with a symmetrized Markov transition kernel. Namely, using a symmetric normalization through conjugation of the forward Markov operator, we introduce the following symmetrized Markov transition kernel~\cite{coifman2005geometric}:
\begin{equation}
  \label{eq:aniso-sym}
  M(\bz, \bs) = \frac{a_{\alpha}(\bz,\bs)}{\sqrt{d_{\alpha}(\bz) d_{\alpha}(\bs)}},
\end{equation}
which preserves the eigenvalues of the forward Markov transfer operator $\operatorname{\mathcal{M}_f}$. Due to the spectral decomposition, the symmetric Markov transition kernel (\req{eq:aniso-sym}) can be written as~\cite{coifman2005geometric}:
\begin{equation}
  \label{eq:decomp}
  M^{(t)}(\bz,\bs) = 
    \sum_{k=1}^{\infty} \lambda_k^{2t} \psi_k(\bz) \psi_k(\bs),
\end{equation}
where $\pset{\psi_k}$ are its eigenvectors, and the corresponding Markov process can be propagated in the auxiliary time by raising the kernel to the power of $t$ as $M^{(t)}$~\cite{coifman2005geometric}.

\subsection{Data-Driven Construction}
\label{sec:dda}
We now construct a data-driven approximation of the limiting diffusion process considered in \rsct{sec:mto}. Taking the anisotropic diffusion constant as $\alpha=1/2$, the discrete approximation of the anisotropic diffusion kernel that models a transition between samples $\bz_k$ and $\bz_l$ in the reduced space is:
\begin{equation}
  \label{eq:data-aniso-kernel}
  a(\bz_k,\bz_l) = \frac{g(\bz_k,\bz_l)}{\sqrt{\varrho(\bz_k)\varrho(\bz_l)}},
\end{equation}
where the function $g(\bz_k,\bz_l)=\exp\ppar*{-{\pnrm{\bz_k-\bz_l}^2/\varepsilon_{kl}}}$ is a Gaussian kernel with sample-dependent scale matrix $\varepsilon_{kl}$, and $\varrho(\bz_k)=\sum_{l} g(\bz_k,\bz_l)$. The sample-dependent scale matrix is introduced to adjust the transition probabilities to model heterogeneous free-energy landscapes. We estimate the sample-dependent scale matrix by adaptively balancing local and global spatial scales~\cite{rydzewski2024learning}:
\begin{equation}
  \label{eq:scale}
  \varepsilon_{kl}(r) = 
    \pnrm{\bz_k - \eta_r(\bz_k)} \cdot 
    \pnrm{\bz_l - \eta_r(\bz_l)},
\end{equation}
where each term is a ball centered at $\bz$ of radius $\eta_r(\bz)>0$. We define this radius by the fraction of the neighborhood size $r\in[0,1]$, allowing us to decide which scale is more relevant. Specifically, the Gaussian kernel describes a local neighborhood around each sample for values $r$ close to 0 (i.e., the nearest neighbors), which correspond to deep and narrow states. For values of $r$ around 1 (the farthest neighbors), it considers more global information, corresponding to shallow and wide states~\cite{rydzewski2024learning}.

\subsection{Spectral Gap}
The dominant eigenvalues of the Markov transition matrix (and, by the connection outlined in \rsct{sec:mto}, to the Fokker--Planck diffusion equation) decay exponentially and are related to the slowest relaxation timescales in the system. They can be identified by associating each eigenvalue with an effective timescale~\cite{bovier2002metastability}, $t_k=-1/\log\lambda_k$. The largest gap in the eigenspectrum is the spectral gap and determines the degree of timescale separation between the slow and fast processes in the system~\cite{rydzewski2023spectral}:
\begin{equation}
  \label{eq:spectral-gap}
  \sigma = \lambda_{k-1} - \lambda_{k},
\end{equation}
and the number of metastable states in the reduced space $k > 0$~\cite{gaveau1996master,gaveau1998theory,bovier2002metastability}. As explained in \rsct{sec:fpd}, to achieve the reduced dynamics that is effectively Markovian, it is crucial to have a gap between neighboring eigenvalues, along with the near degeneracy of the dominant eigenvalue. This condition is essential for the maximal spectral gap at $k$ to lead to the separation into $k$ metastable states. For these reasons, spectral map maximizes the spectral gap in the learning of the reduced representation. 

\subsection{Kinetic Partitioning}
\label{sec:clustering}
The asymptotic rate of convergence of a Markov chain to the equilibrium distribution is determined by the spectral gap of the corresponding Markov transition matrix~\cite{boyd2004fastest,boyd2006convex}. The larger the spectral gap, the faster the Markov chain converges to its equilibrium distribution. Many techniques exploit this result for clustering~\cite{lafon2006diffusion,van2008graph,hummer2015optimal,martini2017variational,kells2019mean}. Such procedures can also be viewed as coarse-graining transition probabilities.

As spectral map maximizes the spectral gap from Markov transition matrices, implementing an algorithm for kinetic partitioning is an easy extension. Namely, by \req{eq:decomp}, we have:
\begin{equation}
  \label{eq:prop}
  M^{(t)}(\bz_k,\bz_l) = \pr\pset[\big]{\bz_{t_0+t}=\bz_l \cond \bz_{t_0}=\bz_k},
\end{equation}
where propagating the Markov chain describing diffusion from $t_0$ by $t$ steps is equivalent to raising the Markov transition matrix to the power of $t$. As the Markov chain converges to the equilibrium distribution, each walk ends in a metastable state.

The resulting metastable states can partly overlap, creating a transition state ensemble comprising paths crossing the overlapping region with similar transition probabilities to sink to either of the states. This definition of transition states is similar to the definition of transition states provided by Hummer~\cite{hummer2004transition}, i.e., transition states can be defined as regions with the highest probability that trajectories passing through them form transition paths between metastable states.

\subsection{Algorithm}
\label{sec:algorithm}
The spectral map framework algorithm is presented in Alg.~\ref{alg:spectral-map}. It employs a neural network as a target mapping (and thus CVs) to map high-dimensional samples to the reduced space. The learning algorithm iterates over epochs and processes batches created from the permuted dataset of samples from a molecular dynamics simulation. \chg{We use the Adam optimizer with a learning rate of $10^{-3}$ (and default parameters) to maximize the spectral gap score (\req{eq:spectral-gap}). For every target mapping, we use linear models, i.e., $\xi_w(\bx) = \sum_{k=1}^n w_k x_k + w_0$, where $w_k$ are learnable parameters. Each target mapping is trained for 100 epochs with data batches consisting of 2000 samples. Following a protocol outlined in~\rref{rydzewski2024learning}, the fraction of neighborhood size used to estimate sample-dependent scale matrices $\varepsilon_{kl}$ is set to $r=0.65$ by determining which value of $r$ corresponds to the largest spectral gap (Fig. S1 in the Supporting Information). After the target mappings converge, we coarse-grained the estimated Markov transition matrices constructed from data batches from the last epoch to obtain a kinetic partitioning of the reduced space. As cluster labels are not assigned uniquely for each batch, we use agglomerative clustering with Ward's minimum variance method to merge clusters. For the case where the reduced space is two-dimensional, we fit a support vector classifier to detect state boundaries. We use PyTorch~\cite{paszke2019pytorch} and scikit-learn~\cite{scikit-learn} for the implementation.}

\begin{algorithm}
  \SetKwInOut{Input}{Input}
  \SetKwInOut{Output}{Output}
  \Input{\chg{Dataset, number of CVs $d$, number of metastable states $k$.}}
  \Output{Target mapping $\xi_{w}(\bx)=\bz$ for CVs with maximal spectral gap.}
  \begin{minipage}{0.9\linewidth}
  \begin{enumerate}
    \item Iterate over training epochs:
      \begin{enumerate}
        \item Map samples to their reduced representation using the target mapping (\req{eq:target-mapping}).
        \item Iterate over data batches:
      \begin{enumerate}
        \item Calculate the symmetric Markov transition matrix $M$ from the anisotropic diffusion kernel (\req{eq:data-aniso-kernel}).
        \item Perform eigendecomposition and estimate the spectral gap as $\sigma=\lambda_{k-1} - \lambda_k$ (\req{eq:spectral-gap}).
        \item Update weights $w$ by maximizing the score given by the spectral gap.
        \item If learning converged, propagate the Markov transition matrix by $M^{(t)}$ to cluster batch samples (\req{eq:prop}).
      \end{enumerate}
    \end{enumerate}
    \item Merge batch clusters to get kinetic partitioning of samples into $k$ metastable states and additional transition states \chg{using an agglomerative clustering algorithm and (optionally) fit a classifier to detect state boundaries in the reduced space}.
  \end{enumerate}
  \end{minipage}
  \caption{Spectral map with kinetic partitioning (coarse-grained diffusion).}
  \label{alg:spectral-map}
\end{algorithm}
\begin{figure*}[t]
  \includegraphics{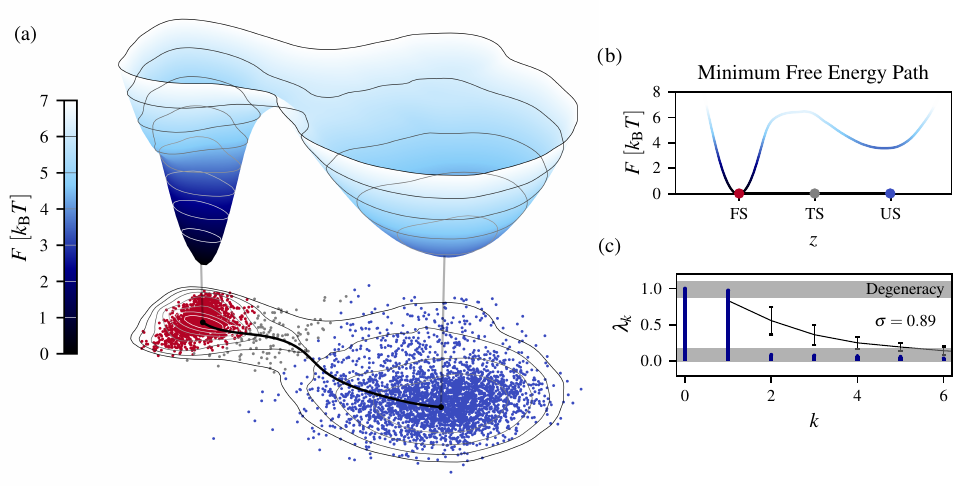}
  \caption{{Spectral map of FiP35.} (a) Free-energy landscape with the folded and unfolded metastable states of FiP35 spanned by two CVs learned by spectral map. The contour lines are placed every 1 $\kT$. The projection with colored samples is shown below the free-energy landscape, where the folded (FS), transition (TS), and unfolded (US) states are shown in red, grey, and blue, respectively. The minimum free-energy path is shown by the black line linking the folded and unfolded states. (b) The minimum free-energy path [corresponding to the black line in (a)] shows the energy barrier between the metastable states of around 5 $\kT$. (c) Eigenspectrum of the Markov transition matrix at the end of the learning procedure showing a large spectral gap of $\sigma=0.89$ between the first and second eigenvalues, i.e., $\sigma=\lambda_1-\lambda_2$. The maximization of the spectral gap results in the degeneracy of the first eigenvalue $\lambda_1$ and the rest of the eigenvalues for close to 0 and thus negligible. The gray line with error bars shows the average and standard deviations of eigenspectra resulting from attempts to maximize the spectral gaps for $k>2$, showing the lack of any significant timescale separation.}
  \label{fig:ww-2d}
\end{figure*}

\section{Results}
\label{sec:results}
In our results, we mainly examine slow CVs learned by spectral map and their ability to accurately describe metastable states and rare transitions between them. As a high-dimensional example, we employ spectral map framework to investigate the reversible folding process of the FiP35 protein in solvent, a member of a WW domain, consisting of two $\beta$ hairpins that form a three-stranded $\beta$ sheet. A 100-$\mu$s unbiased molecular dynamics simulation of FiP35 at its computationally estimated melting temperature of 395 K is obtained from the first trajectory of FiP35 from data provided by D. E. Shaw Research~\cite{shaw2010atomic}. Although this dataset has been analyzed in detail, every such analysis depends on the reaction coordinate used, which makes it suitable to investigate here.

We represent the high-dimensional trajectory by the pairwise Euclidean distances between C$\alpha$ atoms of FiP35, yielding $n=595$ configuration variables (i.e., features) in total, i.e., $\bx=\pset{x_{kl}}_{k>l}^n$ where $x_{kl}$ is the distance between atoms $k$ and $l$. The training dataset consists of 10,000 samples extracted from the trajectory every 10 ns. \chg{To demonstrate the predictive ability of spectral map and, additionally, simplify the analysis, we use linear models as the target mappings. After the target mappings converge, we map the samples from the trajectory (every 200 ps) to construct free-energy landscapes.}

\begin{figure*}[t]
  \includegraphics{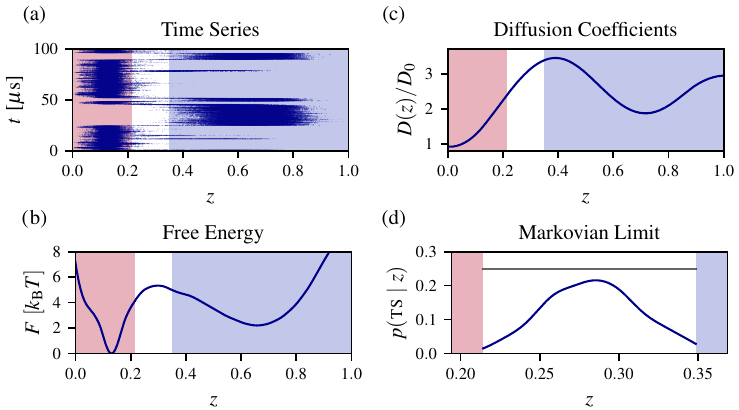}
  \caption{Slow CV $z$ for the FiP35 folding learned by spectral map with kinetic partitioning performed. The folded, unfolded, and transition states are shown in red, blue, and white, respectively. (a) Trajectory $z(t)$ of 100 $\mu$s used for learning. (b) Free-energy profile along $z$ with two metastable states ($\sigma=0.87$) with a barrier of around 5 $\kT$. (c) Coordinate-dependent diffusion coefficients $D(z)/D_0$, where $D_0$ is the diffusion coefficient for the folded state. (d) Probability $p(\textsc{ts} \cond z)$ in the transition state \textsc{ts} with a maximum $p^*=0.22$ that is very close to the Markovian limit of 0.25 for the dynamics in the overdamped regime.}
  \label{fig:ww-1d-markov}
\end{figure*}

\subsection{Timescale Separation}
To assess to what extent we can improve timescale separation in the reduced space, we begin by using spectral map to learn a two-dimensional subspace spanned by CVs (\rfig{fig:ww-2d}a). The learning algorithm determines two metastable states ($k=2$) by optimizing the spectral gap computed through eigendecompositions of Markov transition matrices estimated from data batches. It converges to a maximum spectral gap value of $\sigma=0.89$. This result indicates that the slow CVs achieve a very large timescale separation with two first dominant eigenvalues ($\lambda_0$ and $\lambda_1$) degenerated to one and the rest close to zero. To ensure that the highest timescale separation is indeed for $k=2$ metastable states, we also learn CVs for $k>2$. However, each such learning ends with a negligible timescale separation compared to the result obtained for $k=2$, as indicated by the eigenspectra shown in \rfig{fig:ww-2d}c.

This result shows that the folding of FiP35 is governed by a single dominant slow process related to the reversible transition between the folded and unfolded states (\rfig{fig:ww-2d}a). We observe no indication of distinguishable intermediate or misfolded states, which are ubiquitous in molecular processes. Although seemingly obvious, this conclusion highlights an important aspect of the folding process of FiP35 in our analysis. It possibly indicates the lack of concomitant structurally heterogeneous paths proceeding through any intermediate of misfolded states, which has been previously suggested for FiP35 based on the formation order of the individual hairpins~\cite{noe2009constructing,mori2015dynamic}. However, as spectral map works mainly by unmixing the slow and fast timescales, it is important to note that our conclusions are based on the slow kinetics of the folding of FiP35. This means that if multiple paths coexist on a similar timescale, they can be collapsed in the reduced representation. Nonetheless, our results are consistent with findings presented in \rref{shaw2010atomic}.

\subsection{Free-Energy Landscape}
We calculate the free-energy landscape of folding by mapping the complete trajectory (sampled every 200 ps) to the slow CVs learned by spectral map (\rfig{fig:ww-2d}a). Additionally, we trace out a minimum free-energy path in the reduced space to clearly reveal a free-energy barrier of around 5 $\kT$ separating the folded and unfolded states. The height of the barrier, much higher than the thermal energy $\kT$, confirms that the transition between those metastable states is indeed a slow process. 

It is rather clear from the calculated free-energy landscape that only one slow CV is sufficient to describe the folding process of FiP35, as the second orthogonal CV describes only faster fluctuations in the metastable states occurring on nanosecond timescales (\rfig{fig:ww-2d}a). Therefore, to make the following analysis clearer, we base the following investigation on a single slow CV. To avoid any projection errors from computing ensemble averages, instead of using the slow CV from the two-dimensional reduced space, we run spectral map to learn a new single slow CV. As a result, we obtain a free-energy profile along the resulting one-dimensional CV with every detail of the minimum free-energy path and the spectral gap of $\sigma=0.87$ (\rfig{fig:ww-1d-markov}a), in close agreement with the spectral gap for the two-dimensional reduced space.

Interestingly, the calculated free-energy barrier is higher than previously estimated for structural reaction coordinates~\cite{liu2008experimental}. A vanishingly small barrier of around 3 $\kT$ has suggested that FiP35 follows a downhill folding scenario, which is the reason for its folding on microsecond timescales. Our results show rather that the significant timescale separation represented by the slow CV has a slight tendency to deviate from downhill folding to a two-state process~\cite{kubelka2004protein}.

\subsection{Transition Path Ensemble}
As shown in the case of both two-dimensional (\rfig{fig:ww-2d}a) and one-dimensional (\rfig{fig:ww-1d-markov}) reduced space, we observe two main ``sink'' regions corresponding to the folded and unfolded states. By examining the coarse-grained transition probabilities of the Markov matrices, we identify a partially overlapping region. The main reason for this overlap is that the trajectory in this transition region has similar transition probabilities into either folded or unfolded states. We use this unstable transition region to define a transition state ensemble. We can see that the location of the transition state is also consistent with the curvature of the free-energy profile as it symmetrically encompasses a dynamical bottleneck in the form of the free-energy barrier (\rfig{fig:ww-1d-markov}a). \chg{Additionally, we verify the clustering in the two-dimensional reduced space by fitting a support vector classifier to compute state boundaries (Fig. S2 in the Supporting Information).}

Using the defined transition region, we estimate mean first-passage times (MFPTs) directly from the trajectory by counting transitions between the folded and unfolded states of FiP35. For this, we calculate MFPTs for several values of lag times. The resulting MFPTs reach a constant value for a lag time of around 0.1 $\mu$s, yielding $\sim$13 $\mu$s for the transition from the folded to the unfolded states and $\sim$6 $\mu$s for the reverse transition. As expected in such calculations, due to the limited number of transitions between the states, the 0.95 confidence intervals for the MFPTs estimated by bootstrapping are high (Fig. S3 in the Supporting Information). Despite the large error margins, our results are in relative agreement with experiments~\cite{liu2008experimental}.

\subsection{Markovianity}
We expect that the slow CVs found by spectral map should exhibit Markovian behavior~\cite{rydzewski2024learning}. As timescale separation increases during the learning process, long memory effects in the reduced dynamics are minimized~\cite{berezhkovskii2011time,guttenberg2013minimizing,rydzewski2023spectral,rydzewski2024learning}. This assumption can be confirmed by employing a simple Bayesian test proposed by Berezhkovskii and Makarov~\cite{berezhkovskii2018single}. Although the test is derived using a committor, it can be applied to any one-dimensional coordinate. Namely, for a transition region \textsc{ts} bounded in $[a, b]$ along a reaction coordinate $z$, the following equality holds when the reduced dynamics exhibits a Markovian behavior:
\begin{equation}
  \label{eq:sasha}
  p^* \equiv \max_{a \leq z \leq b} p(\textsc{ts} \cond z) = 1/4,
\end{equation}
where $p(\textsc{ts} \cond z)$, estimated from Bayes' theorem~\cite{hummer2004transition,berezhkovskii2018single}, is the probability for a trajectory to sample a transition state, given that the system is in $z$. In contrast, for non-Markovian dynamics, we have $p^* < 1/4$. The equality in \req{eq:sasha} holds for dynamics in the overdamped regime and (as a result of the detailed balance) does not depend on the direction of the transition.

Using the transition region determined by the kinetic partitioning algorithm, we perform the test for Markovianity. We can see in \rfig{fig:ww-1d-markov}c that the slow CV learned by spectral map closely approaches the ideal Markovian limit for memoryless stochastic processes (i.e., $p^* = 0.22$). The Markovianity of the slow CV is also consistent with the assumptions that we made in \rsct{sec:framework}, where we choose to work with the Fokker-Planck diffusion in the overdamped regime. This is a noteworthy observation as it demonstrates the possibility of constructing a single reaction coordinate that is easier to understand due to its dominant Markovian characteristics. This is in contrast to non-Markovian variables that necessitate including complicated memory effects for a process such as protein folding. Moreover, previous studies on proteins of sizes comparable to that of FiP35 have shown that folding observed along structure-based coordinates, such as the root-mean-square deviation from the native structure, reveal strong non-Markovian effects and anomalous diffusion~\cite{satija2017transition}.

\subsection{Coordinate-Dependent Diffusion Coefficients}
Coordinate-dependent diffusion coefficients $D(z)$ are important attributes of the reduced dynamics~\cite{zwanzig1992diffusion,berezhkovskii2011time} and can affect a free-energy landscape by shifting transition states and barrier height of protein folding~\cite{chahine2007configuration}. However, this effect highly depends on a coordinate chosen to quantify the reduced dynamics~\cite{rhee2005one,krivov2008diffusive,best2010coordinate,hinczewski2010diffusivity}. Using a deficient CV that is affected significantly by the diffusion coefficients can cause rate-limiting and kinetic bottlenecks. To address this issue in FiP35, we need to check whether learning a Markovian CV that accounts for the slowest timescale in the system can reduce the effect of diffusion coefficients on the dynamics in a free-energy landscape.

To this aim, we calculate diffusion coefficients through linear regression to the fluctuations of the slow CV in the short time limit. To assess how $D(z)$ affects the free-energy landscape $F$, we define a diffusion-dependent free-energy as:
\begin{equation}
  F_D(z) = F(z) - \frac{1}{\beta}\log \ppar*{\frac{D(z)}{D_0}},
\end{equation}
where $D_0$ is the value of the diffusion coefficient for the folded state (\rfig{fig:ww-1d-markov}c). Our results agree with our expectation that the metastable states are less affected than the transition region near the barrier. However, this effect is small, given that the folding rates depend linearly on the diffusion coefficients but exponentially on the barrier height. Here, the average discrepancy from the free-energy profile $F(z)$ is below $\kT$ (Fig. S4 in the Supporting Information), indicating that the learned slow CV represents the long-timescale dynamics accurately and is not affected by additional friction coming from the remaining fast degrees of freedom~\cite{berezhkovskii2011time}.

\subsection{Structural Descriptors}
Finally, we compare the slow CV learned by spectral map and its ability to distinguish between timescales with commonly used structural descriptors to quantify folding processes. We estimate spectral gaps for the fraction of native contacts and the features used to construct the reduced space, i.e., the pairwise distances between C$\alpha$ atoms of FiP35, $\bx=\pset{x_{kl}}_{k>l}^n$ where $x_{kl}$ is the distance between atoms $k$ and $l$. We calculate the fraction of native contacts as~\cite{best2013native}: 
\begin{equation}
  \label{eq:frac}
  q(\bx) = \frac{1}{n} \sum_{k>l}^n \ppar*{1+\e^{\alpha(x_{kl}-\gamma x^0_{kl})}}^{-1},  
\end{equation}
where $n$ is the total number of native contacts, $x^0_{kl}$ is the distance between atoms $k$ and $l$ in the folded structure, $\alpha=50$ nm$^{-1}$, and $\gamma=1.5$. In \req{eq:frac}, we include only pairwise distances that are less than 0.8 nm and the residues they belong to have a difference in sequence position greater than three.

\begin{figure*}[t]
  \includegraphics{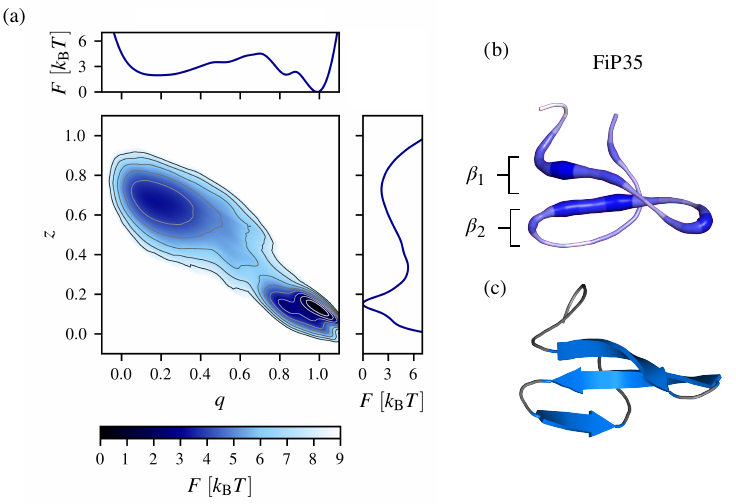}
  \caption{(a) Free-energy landscape $F(\bz)$ as a function of $\bz=(z,q)$, where $z$ is the learned slow CV (the corresponding spectral gap $\sigma=0.87$) and $q$ is the fraction of native contacts ($\sigma=0.72$). (b) FiP35 residues important for its transitions on the longer timescales shown on a randomly selected conformation from the folded state (shown in blue). The three strands of FiP35 form the $\beta_1$ and $\beta_2$ sheets are depicted. FiP35 residues are colored according to their relative importance calculated using spectral gaps of pairwise Euclidean distances: $r_k = \sum_l \sigma_{kl}$, where $\sigma_{kl}$ denotes the spectral gap of the distance between residues $k$ and $l$. (c) Crystallographic structure of FiP35 shown for comparison~\cite{jager2006structure}.}
  \label{fig:ww}
\end{figure*}

The fraction of native contacts has successfully been used for many proteins~\cite{best2013native}. As expected, the fraction of native contacts with the spectral gap of $\sigma=0.72$ is the closest to the slow CV. Both profiles are qualitatively similar (\rfig{fig:ww}a); however, the free-energy barrier along the fraction of native contacts is slightly lower compared to the slow CV. This is consistent with the fact that the larger spectral gap induces a better separation of metastable states~\cite{rydzewski2023spectral,rydzewski2024learning}. The profile $F(q)$ has a small indentation at $q\sim 0.8$. However, the barrier separating this space from the folded state is too small to create an intermediate state. 

Spectral gaps of the pairwise distances are considerably lower in comparison to the spectral gap of the slow CV and a free-energy barrier between the metastable states $< 5$ $\kT$ (Fig. S5 in the Supporting Information), indicating that using a single pairwise distance as a reaction coordinate does not results in an optimal separation of timescales. However, several features have spectral gaps $\sim$0.65 and possibly contribute to the level of timescale separation in the reduced dynamics (Fig. S6 in the Supporting Information). These features correspond to two groups of residues that make up the majority of interactions important for the folding process of FiP35: Pro5--Trp8 (the first $\beta$ strand) and Met12--Asp15 (the tip of the first $\beta$ hairpin). Additionally, the distance between Phe21 and Ser28 has a large spectral gap (between the second and the third $\beta$ strands). Belonging to these regions are residues identified as important for the folding, and some of them have been selected for mutational analysis. Notably, the mutation of Phe21 to Leu in \rref{shaw2010atomic} has resulted in a free-energy change of 2.4 kcal/mol.

By mapping all the features back to the FiP35 residues and calculating their relative importance (\rfig{fig:ww}b), we can show the structural parts of FiP35 that are important for its long-time folding dynamics. Namely, the main regions involved in this process are the strands constituting the first hairpin and the loop between these strands, indicating that the first hairpin plays a larger role in the folding compared to the second hairpin (\rfig{fig:ww}b and c). This is in agreement with other studies showing that the rate-limiting step in the folding reaction involves the formation of the first hairpin~\cite{deechongkit2004context}. Summarizing, we can use spectral map also to identify the structural parts of FiP35 contributing to slow structural signal propagation during the folding. We think the above protocol can be used to select residues for mutational analysis to guide experiments.

\section{Conclusions}
\chg{
We have further developed the framework of spectral map, an unsupervised statistical learning technique. Spectral map learns slow CVs by maximizing a spectral gap between slow and fast eigenvalues of a Markov transition matrix. By introducing a simple improvement to the learning algorithm, we have shown that transition state ensembles can also be learned. 

This has helped us to show that even complex molecular processes such as protein folding, often modeled by employing a generalized Langevin equation with a memory kernel, can be reduced to dynamics with diffusive and Markovian characteristics that can be fully described by a free-energy landscape along a single slow CV and the associated diffusion coefficients. This result is important as it is more common for the reduced dynamics along a reaction coordinate to exhibit non-Markovian effects, where free-energy barriers can be not especially indicative as correlation described by the memory kernel leads to subdiffusive dynamics and compensates low barriers to preserve reaction kinetics. We have also shown that the free energy is not significantly affected by the coordinate-dependent diffusion coefficients. This strongly suggests that the slow CV learned by spectral map can be used as a reaction coordinate to quantify physical characteristics of protein folding, such as dominant metastable and transition states~\cite{socci1996diffusive,best2006diffusive}.

We have demonstrated that the folding process of FiP35 does not proceed hierarchically (i.e., not through any intermediate or misfolded states) in the reduced representation learned by spectral map. This is evident from the absence of multiple pathways connecting the folded and unfolded states. It should be noted, however, that multiple structural paths to transition between the folded and unfolded states of FiP35 can be present, but they can coexist on a similar timescale. As spectral map unmixes the slow and fast timescales, this can lead to merging these paths in the reduced representation.

The formalism described here was initially introduced for anisotropic diffusion maps that apply it to the configuration space to perform eigendecomposition and approximate the reduced space by the eigenvectors of the Markov transition kernel based on the spectral gap~\cite{coifman2005geometric,coifman2006diffusion}. In spectral map, the eigendecomposition is employed in the reduced space to maximize the spectral gap iteratively. It is also important that in spectral map, the number of CVs used to describe the reduced dynamics is not indicated by the spectral gap position. This is in contrast to methods that use eigenvectors to parametrize the reduced space.

Spectral map can currently be used to learn from unbiased molecular dynamics simulations, as it does not incorporate a reweighting algorithm into its framework. There are many such methods for reweighting Markov transition matrices constructed from trajectories sampled by enhanced sampling techniques~\cite{donati2017girsanov,donati2018girsanov,kieninger2020dynamical,donati2022review,bolhuis2023optimizing,shmilovich2023girsanov}. However, we think that implementing reweighting for anisotropic diffusion kernels~\cite{zhang2018unfolding,rydzewski2021multiscale,rydzewski2022reweighted,rydzewski2023selecting,rydzewski2023manifold}, as we previously introduced, is the easiest way to extend spectral map to learn from biased data. We plan to publish our results on this topic in the near future.

Overall, we have shown that spectral map can be used to learn slow reaction coordinates for molecular processes with multiple timescales and enable us to understand the underlying physics. Together with our previous work~\cite{rydzewski2023spectral,rydzewski2024learning}, our results have indicated that spectral map is a promising technique and merits further development.
}

\begin{suppinfo}
  Supporting Information is available free of charge at \url{https://pubs.acs.org/}.
  \begin{itemize}
    \item Comment on the FiP35 dataset.
    \item Additional figures: estimation of the fraction of neighborhood size; support vector classifier trained based on kinetic partitioning; mean first-passage times for transitions between the folded and unfolded states; coordinate-dependent diffusion coefficients; free-energy profiles and eigenspectra for pairwise distances; dependence of the spectral gap on Pearson correlation coefficients.
  \end{itemize}
  \end{suppinfo}

\section*{Acknowledgements}
We thank A. Berezhkovskii for valuable discussions. We acknowledge support from the Polish Science Foundation (START), the Ministry of Science and Higher Education in Poland, the Japan Society for the Promotion of Science (JSPS), and the National Science Center in Poland (Sonata 2021/43/D/ST4/00920, ``Statistical Learning of Slow Collective Variables from Atomistic Simulations''). D. E. Shaw Research is acknowledged for providing the FiP35 dataset.

\bibliography{main.bib}

\end{document}